# The Statistical Distribution of quantum particles where any of them can be a fermion with probability [P] or a boson with probability [1-p].

## Ahmad Abu Taleb[*]

**Department of Mathematics, Faculty of Science, Mansoura University, Egypt.**

**Abstract:**

Despite the obvious difference between fermions and bosons in their physical properties and statistical distributions, but we have to ask the following question. What is the form of statistical distribution for a system of quantum particles where any of them can be a fermion with probability [P] and can be a boson with probability [1-P]?.
In this paper this question is solved.

**Introduction:**

The well-known distribution which describe how the number of indistinguishable fermions or bosons in different energy states varied with the energy of those states are Fermi-Dirac distribution and Bose-Einstein distribution. [3].

In this short paper the new is that we deduce the form of the statistical distribution of a system of indistinguishable particles where any of them can be a fermion with probability [P] and can be a boson with probability [1-P].

As expected we find that both of Fermi-Dirac distribution and Bose-Einstein distribution seems to be limit cases of this distribution.

**Analysis:**

Suppose that we have a number of energy levels, labeled by index (i), each level having energy ($\varepsilon_i$) and containing a total of ni particles. Suppose each level containing (gi) distnistic sublevels, all of which have the same energy and which are distinguishable.

---

[*] *E-mail address*; eng.ahamadadel@yahoo.com

If those particles were fermions then the number of ways of distributing ni indistinguishable fermions among the gi sublevels of an energy level is

$$w_f(n_i, g_i) = \frac{g_i!}{n_i!(g_i-n_i)!} \qquad (1)$$

and the total number of ways that a set of occupation numbers ni can be realized is the product of the ways that each individual energy level can be populated

$$W_f = \prod_i \frac{g_i!}{n_i!(g_i-n_i)!} \qquad (2)$$

On the other hand if those particles were bosons then the number of ways of distributing ni indistinguishable bosons among the gi sublevels of an energy level is

$$w_b(n_i, g_i) = \frac{(n_i+g_i-1)!}{n_i!(g_i-1)!} \qquad (3)$$

and the total number of ways that a set of occupation number ni is

$$W_b = \prod_i \frac{(n_i+g_i-1)!}{n_i!(g_i-1)!} \qquad (4)$$

If we assume that each state has a high degeneracy, i.e. gi >> 1, then we can make the approximation

$$w_b(n_i, g_i) = \frac{(n_i+g_i)!}{n_i! g_i!} \qquad (5)$$

And

$$W_b = \prod_i \frac{(n_i+g_i)!}{n_i! g_i!} \qquad (6)$$

We want to find the set of ni for which wf is maximized for the case of fermions and for wb is maximized for the case of bosons, subject to the constraint that there be a fixed number of particles and a fixed energy.

We constrain our solution using Lagrange multipliers forming the function for fermions

$$f_f(n_i) = \ln(W_f) + \alpha(N - \Sigma_i n_i) + \beta(E - \Sigma_i n_i \varepsilon_i) \quad (7)$$

and forming the following function for bosons

$$f_b(n_i) = \ln(W_b) + \alpha(N - \Sigma_i n_i) + \beta(E - \Sigma_i n_i \varepsilon_i) \quad (8)$$

Where $N$ is the total number of particles and E is the total energy.

For both of $F_f(n_i)$ and $F_b(n_i)$, using Sterling approximation for the factorials, taking the derivative with respect to ni, setting the result to zero, and solving for ni yields the Fermi-Dirac distribution

$$n_i = \frac{g_i}{e^{\alpha + \beta \varepsilon_i} + 1} \quad \text{for the function } f_f(n_i) \quad (9)$$

it can be shown thermodynamically that $\beta = 1/KT$ and $\alpha = \mu/KT$, [2], where ($\alpha$) is the chemical potential, $K$ is Boltzmann's constant and $T$ is the temperature, so finally we have: Fermi - Dirac distribution

$$n_i = \frac{g_i}{e^{(\varepsilon_i - \mu)/kt} + 1} \quad (10)$$

and Bose-Einstein distribution

$$n_i = \frac{g_i}{e^{(\varepsilon_i - \mu)/kt} - 1} \quad (11)$$

Now the question is what is the form of the distribution if any of the particles can be a fermion with probability (P) or can be a boson with probability (1-P)? This is equivalent to say that if we have ni particles then we have $(Pn_i)$ fermions and $(1-P)n_i$ bosons.[1]

Therefore the number ways of distributing $(Pn_i)$ fermions and (1-P)ni bosons among the gi sublevels of an energy level is

$$w_a(n_i, g_i) = \frac{g_i!}{pn_i!(g_i - pn_i)!} * \frac{(g_i + (1-p)n_i - 1)!}{((1-p)n_i)!(g_i - 1)!} \quad (12)$$

and for $g_i \gg 1$ we have

$$w_a(n_i, g_i) = \frac{g_i!}{pn_i!(g_i-pn_i)!} * \frac{(g_i+(1-p)n_i)!}{((1-p)n_i)!g_i!} \quad (13)$$

and the total number of ways that a set of occupation number (ni) is

$$W_a = \prod_i \frac{g_i!}{pn_i!(g_i-pn_i)!} * \frac{(g_i+(1-p)n_i)!}{((1-p)n_i)!g_i!} \quad (14)$$

We have

$$\lim_{p \to 1^-} W_a = W_f \quad (15)$$

and

$$\lim_{p \to 0^+} W_a = W_b \quad (16)$$

We form the function

$$f_a(n_i) = \ln(W_a) + \alpha(N - \Sigma_i n_i) + \beta(E - \Sigma_i n_i \epsilon_i) \quad (17)$$

and by the same procedure of finding Fermi-Dirac distribution and Bose-Einstein distribution we find

$$\left(\frac{g_i}{(1-p)n_i}+1\right)^{1-p} \left(\frac{g_i}{pn_i}-1\right)^p = e^{(\epsilon_i-\mu)/kt} \quad (18)$$

but we have from calculus

$$\lim_{p \to 1^-} \left(\frac{g_i}{(1-p)n_i}+1\right)^{1-p} = 1 \quad \text{(limit from the lift)}$$

and

$$\lim_{p \to 0^+} \left(\frac{g_i}{pn_i}-1\right)^p = 1 \quad \text{(limit from the right)}$$

Then if we taking the limit of (18) as P →1⁻ we have

$$\left(\frac{g_i}{n_i}-1\right) = e^{(\epsilon_i-\mu)/kt} \quad (19)$$

or

$$n_i = \frac{g_i}{e^{(\epsilon_i-\mu)/kt}+1} \quad (19\ A) \quad \text{(Fermi- Dirac distribution)}$$

and taking the limit of (18) as P $\to 0^+$

we have

$$\left(\frac{g_i}{n_i}+1\right) = e^{(\epsilon_i-\mu)/kt} \qquad (20)$$

or

$$n_i = \frac{g_i}{e^{(\epsilon_i-\mu)/kt}-1} \qquad (20\ A)\ (\text{Bose - Einstein distribution})$$

Then we find that Fermi-Dirac distribution is a limit case of distribution (18) as P $\to 1^-$ and Bose-Einstein distribution is a limit case of distribution (18) as P $\to 0^-$.